\documentclass[aps,twocolumn,superscriptaddress,showpacs,prb]{revtex4}
\usepackage{graphicx,amsmath}
\newlength{\figwidth}
\begin{document}
\title{Spatial fluctuations of spin and orbital 
in two-orbital Hubbard model}
\author{Tomoko Kita}
\affiliation{
Department of Applied Physics, Osaka University, 
Suita, Osaka 565-0871, Japan
}
\author{Takuma Ohashi}
\affiliation{
Department of Physics, Osaka University, 
Toyonaka, Osaka 560-0043, Japan
}
\author{Sei-ichiro Suga}
\affiliation{
Department of Applied Physics, Osaka University, 
Suita, Osaka 565-0871, Japan
}
\date{\today}
\begin{abstract}
We investigate the quasiparticle dynamics in the two-orbital Hubbard model 
on the square lattice at quarter filling by means of 
the cellular dynamical mean field theory. 
We show that the Fermi-liquid state is stabilized up to 
the large Hubbard interactions 
in the symmetric case without the Hund's coupling, 
and find the heavy quasiparticles around 
the metal-insulator boundary. 
It is elucidated that the Hund's coupling enhances 
the antiferro-orbital correlations, 
which give rise to the pseudo gap behavior 
in the single-particle excitations. 
We also find the nonmonotonic temperature dependence 
in the quasiparticle dynamics for 
intermediate strength of the Hund's coupling, 
and clarify that it is caused by the competition between 
the Fermi-liquid formation and the antiferro-orbital fluctuations. 
\end{abstract}
\pacs{
71.30.+h 
71.10.Fd 
71.27.+a 
} 
\maketitle
\section{Introduction}
Strongly correlated electron systems with orbital degrees of freedom 
have attracted much attention. 
Typical examples are the manganite 
$\mathrm{La}_{1-x}\mathrm{Sr}_x\mathrm{MnO}_3$ \cite{tokura94}, 
the ruthenate $\mathrm{Sr}_2\mathrm{RuO}_4$ \cite{maeno94}, and 
the vanadate $\mathrm{LiV}_2\mathrm{O}_4$ \cite{kondo97}, 
where striking phenomena such as colossal magnetoregistance, 
triplet superconductivity, and heavy fermion behavior 
have been observed. 
The importance of orbital degrees of freedom has been 
suggested also in unconventional superconductors, 
such as 
$\mathrm{Na}_x\mathrm{CoO}_2 \cdot y\mathrm{H}_2\mathrm{O}$ \cite{takada03}
and the newly discovered iron-based superconductor 
$\mathrm{LaFeAsO}_{1-x}\mathrm{F}_x$ \cite{kamihara08}. 

These interesting experimental findings have stimulated a
number of theoretical works on electron correlations 
in the multiorbital systems. 
Among them, 
the dynamical mean field theory (DMFT) \cite{georges96} 
has successfully applied to the multiorbital systems 
\cite{kotliar96,rozenberg97,momoi98,imai01,koga04,
koga05,inaba05,inaba07,arita07,sakai06,sakai07}. 
The long standing issue of itinerant ferromagnetism
has been investigated by DMFT \cite{momoi98,sakai07}, 
and it has been clarified that 
not only the lattice structure 
but also orbital fluctuations 
under strong influence of the Hund's coupling
is important for the realization of ferromagnetism. 
The Mott transition in the multiorbital systems 
has been also studied extensively by means of DMFT. 
One of important conclusions of DMFT is that 
orbital degeneracy strongly enhances fluctuations of 
spin, charge and orbital, which 
stabilize the renormalized metallic state \cite{koga05}. 
The observation of the orbital-selective Mott transition 
in $\mathrm{Ca}_{2-x}\mathrm{Sr}_x\mathrm{RuO}_4$ 
\cite{nakatsuji03} has further stimulated studies on 
the metal-insulator transition in the multiorbital systems. 
The DMFT studies on 
the two-orbital Hubbard model with different bandwidths 
have clarified that orbital fluctuations play a key role 
in the orbital-selective Mott transition 
\cite{koga04,koga05,inaba05,inaba07}. 
However, the effects of the spatial correlations 
on the Mott transition in the multiorbital systems
have not yet been well understood, 
because the DMFT does not treat spatially extended correlations. 
Therefore, 
it is desirable to investigate 
how the spatial fluctuations of spin and orbital 
affect quasiparticle dynamics around the Mott transition 
in the multiorbitals system, 
using a different appropriate method. 

In this paper, we study the effects of 
spatial fluctuations of spin and orbital 
on the quasiparticle dynamics 
in the two-orbital Hubbard model at quarter filling,
using the cellular dynamical mean field theory (CDMFT) \cite{kotliar01} 
combined with the noncrossing approximation (NCA) 
\cite{bickers87,stanescu06}. 
In the absence of the Hund's coupling, 
we show that the metallic state 
persists up to large Coulomb interactions. 
This gives rise to the heavy quasiparticle behavior 
around the metal-insulator boundary. 
It is clarified that the Hund's coupling 
strongly enhances the antiferro-orbital (AFO) fluctuations, 
which suppress the quasiparticle formation. 
The strong Hund's coupling induces 
the pseudo gap in the single-particle spectra. 
In the intermediate regime of the Hund's coupling, 
we find more striking behavior, namely, 
a novel nonmonotonic temperature dependence 
of the single-particle excitations: 
the heavy quasiparticles once develop with lowering temperatures, 
and then they are suppressed at much lower temperatures. 
It is elucidated that 
this nonmonotonic temperature dependence of 
the single-particle excitations is 
caused by the competition between 
the quasiparticle formation and the AFO correlations. 

\section{Model and Method}
We consider the two-orbital Hubbard model 
on the two-dimensional square lattice, 
\begin{align}
H &= - t \sum _{\langle i,j \rangle \alpha \sigma} 
c_{i \alpha \sigma}^\dag c_{j \alpha \sigma} 
+ U \sum _{i \alpha} n_{i \alpha \uparrow }n_{i \alpha \downarrow }
\nonumber \\
&+ \sum _{i \sigma \sigma'} (U'-\delta_{\sigma \sigma'}J) 
n_{i 1 \sigma}n_{i 2 \sigma' } 
\nonumber \\
&- J \sum _{i} \left( 
c_{i 1 \uparrow}^\dag c_{i 1 \downarrow} 
c_{i 2 \downarrow}^\dag c_{i 2 \uparrow} + 
c_{i 1 \uparrow}^\dag c_{i 1 \downarrow}^\dag 
c_{i 2 \uparrow} c_{i 2 \downarrow} + 
\mathrm{h.c}.\right ), 
\label{eqn:model}
\end{align}
where $c_{i \alpha \sigma}^{(\dag)}$ is an annihilation (creation) 
operator with spin $\sigma$ ($=\uparrow,\downarrow$) 
and orbital $\alpha$ ($=1,2$) at the $i$th site, 
and $n_{i \alpha \sigma }=c_{i \alpha \sigma}^\dag c_{i \alpha \sigma}$ 
is the number operator. 
Here, 
$t$ denotes the nearest-neighbor hopping integral, 
$U (U')$ the intra-orbital (inter-orbital) Coulomb interaction, and 
$J$ the Hund's coupling including the spin-flip and pair-hopping terms. 
We impose the condition $U=U'+2J$ due to the symmetry requirement. 

\begin{figure}[tb]
\begin{center}
\includegraphics[clip,width=\figwidth]{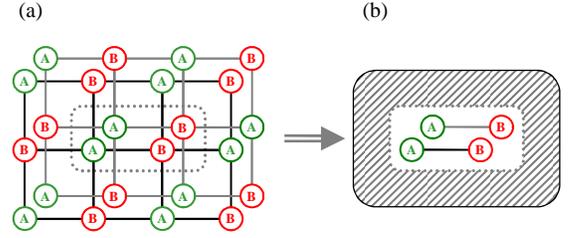}
\caption{
(Color online) 
(a) Sketch of the two-orbital Hubbard model 
on the square lattice  and 
(b) the effective cluster model 
using two-site/two-orbital cluster CDMFT. }
\label{fig:model}
\end{center}
\end{figure}

In order to investigate the effects of 
spatially extended spin and orbital fluctuations 
around the Mott transition, 
we employ CDMFT, a cluster extension of DMFT. 
In CDMFT, the original lattice is regarded 
as a superlattice consisting of clusters, which 
is then mapped onto an effective cluster model 
via a DMFT-like procedure. 
We use a two-site cluster model 
coupled to the self-consistently determined medium. 
We note that the tiling of the original lattice in 
two-site clusters is not unique. 
Here, we define each site in the two-site cluster 
as the site $A$ and $B$. 
In order to incorporate the antiferromagnetic (AFM) and AFO correlations, 
we choose a tiling configuration 
as shown in Fig. \ref{fig:model}, 
where the site $A$ is next to the site $B$. 
Using the above method, 
we can treat the short-range correlations of spin and orbital 
within the two-site cluster. 
The effective cluster Hamiltonian is obtained as, 
\begin{align}
H_\mathrm{eff} &=
\sum _{ij \alpha \sigma} t_{ij} 
c_{i \alpha \sigma}^\dag c_{j \alpha \sigma}
- \mu \sum_{i \alpha \sigma}
n_{i \alpha \sigma }
\nonumber \\
&+ U \sum _{i \alpha}
n_{i \alpha \uparrow }n_{i \alpha \downarrow }
+ \sum _{i \sigma \sigma'} (U'-\delta_{\sigma \sigma'}J)
n_{i 1 \sigma}n_{i 2 \sigma' }
\nonumber \\
&- J \sum _{i } \left(
c_{i 1 \uparrow}^\dag   c_{i 1 \downarrow}
c_{i 2 \downarrow}^\dag c_{i 2 \uparrow} +
c_{i 1 \uparrow}^\dag   c_{i 1 \downarrow}^\dag
c_{i 2 \uparrow}        c_{i 2 \downarrow} +
\mathrm{h.c}.\right )
\nonumber \\
&+\sum _{k} \sum_{\alpha \sigma} \varepsilon_{k \alpha \sigma} 
a_{k \alpha \sigma}^\dag a_{k \alpha \sigma}
\nonumber \\
&+ \sum _{k} \sum_{i \alpha \sigma } \left( 
V_{k i \alpha \sigma}
a_{k \alpha \sigma}^\dag c_{i \alpha \sigma} +
\mathrm{h.c}. 
\right), 
\label{eqn:eff}
\end{align}
where $\mu$ is the chemical potential and $t_{ij}$ is 
the hopping matrix elements in the two-site cluster, 
respectively. 
An effective bath fermions with the energy $\varepsilon_{k \alpha \sigma}$ 
are created by $a_{k \alpha \sigma}^\dag$ 
and coupled to electrons in cluster sites via $V_{k i \alpha \sigma}$. 
Here, $i$ corresponds to the cluster indices ($i=A, B$), 
$\alpha$ is the orbital index ($\alpha= 1, 2$), 
$\sigma$ is the spin index ($\sigma=\uparrow, \downarrow$), 
and $k$ labels infinite bath degrees of freedom. 
In the effective model (\ref{eqn:eff}), 
we calculate the cluster Green's function 
$\hat{G}_{\alpha \sigma}$ by means of NCA as,  
\begin{align}
\hat{G}_{\alpha \sigma} (\omega) = \left[
(\omega+\mu)\hat{1}-\hat{t} 
-\hat{\Gamma}_{\alpha\sigma}(\omega)-\hat{\Sigma}_{\alpha\sigma}(\omega)
\right]^{-1}, 
\end{align}
where 
$\hat{\Sigma}_{\alpha\sigma}$ is the self-energy. 
Here, $\hat{O}$ denotes $2\times2$ matrix for fixed $\alpha$, $\sigma$. 
The effective medium is described by the hybridization function 
$\Gamma_{ij \alpha \sigma}(\omega)= 
\sum_k V_{k i \alpha \sigma}V_{k j \alpha \sigma}^*/
(\omega-\varepsilon_{k \alpha \sigma})$. 
In terms of the self-energy $\hat{\Sigma}_{\alpha\sigma}$, 
the hybridization function is recomputed by, 
\begin{align}
&\hat{\Gamma}_{\alpha \sigma} \left( \omega \right) 
= \left( \omega + \mu \right)\hat{1} - \hat{t}
 - \hat{\Sigma}_{\alpha \sigma} \left( \omega \right) \nonumber \\
&-\left[ 
    \frac{1}{N}\sum_\mathbf{K} 
    \frac{1}{\left( \omega + \mu \right)\hat{1} - \hat{t}\left( \mathbf{K} \right) 
    - \hat{\Sigma}_{\alpha \sigma} \left( \omega \right) }
  \right]^{-1}, 
\label{eqn:cdmft}
\end{align}
where the summation of $\mathbf{K}$ is taken over the reduced Brillouin zone 
of the superlattice. 
Here, $\hat{t} \left( \mathbf{K} \right)$ is the 
Fourier-transformed hopping matrix for the superlattice. 
This procedure is iterated until numerical convergence is reached. 

In DMFT or the cluster extensions of DMFT, 
a solver of the effective Hamiltonian 
(\ref{eqn:eff}) is important. 
The numerically exact solvers are 
the numerical renormalization group (NRG) \cite{wilson75,bulla08}, 
density matrix renormalization group (DMRG) \cite{white92} and 
quantum Monte Carlo (QMC) method \cite{hirsch86,werner06}, 
which are computationally expensive. 
As an approximate method, the exact diagonalization (ED), 
iterative perturbation theory (IPT), NCA, etc. have been used \cite{georges96}. 
NRG \cite{bulla99,bulla08} and DMRG \cite{garcia04,nishimoto04,karski05} 
have been successfully applied to single-site DMFT but 
it is difficult to apply them to cluster-DMFT \cite{maier05} or multiorbital systems. 
In cluster-DMFT combined with Hirsch-Fye QMC 
\cite{hirsch86,maier05,moukouri01,parcollet04,sun05,
ohashi06,ohashi08,sakai09}, 
it is known that the sign problem becomes serious due to 
the Hund's coupling or the pair hopping 
in the multiorbital system \cite{sakai06,sakai07}. 
In CDMFT + ED \cite{civelli05,kyung06,zhang07}, the finite-size errors 
in the effective model (finite $k$ points in (\ref{eqn:eff})) 
become larger in the two-band model than the single-band model 
due to the memory limitation. 
IPT, which is the second order perturbation expansion in $U$, 
gives a good approximation in the single-band 
Hubbard model or periodic Anderson model 
in the infinite dimensions \cite{georges96}, 
but it is qualitatively incorrect for the multiorbital models or cluster-DMFT. 
On the other hand, in NCA, 
neither finite-size error nor sign problem arises. 
It is known that 
DMFT + NCA works well not only in the single-band models but also 
in the multiorbital models \cite{imai01}. 
The cluster-DMFT combined with the NCA 
has also been successfully applied to the single-band Hubbard model 
\cite{stanescu06,maier05,imai02,maier00prl,maier00}, 
and $t$-$J$ model \cite{haule07,maier08}. 
In CDMFT + NCA, 
we do not encounter much more serious problems 
in the two-orbital Hubbard model than in the single band model. 
Therefore, we employ NCA as a solver of 
the effective model (\ref{eqn:eff}). 
NCA is a perturbation expansion around the molecular limit 
and computationally inexpensive. 
In general, it gives rise to artificial non-Fermi liquid (NFL) 
properties at lower temperatures than the Fermi liquid (FL) 
coherence temperature \cite{bickers87}. 
In the following discussion, 
we restrict our analysis to the temperature region 
around and higher than the FL coherence temperature.

\section{Results}

\begin{figure}[tb]
\begin{center}
\includegraphics[clip,width=\figwidth]{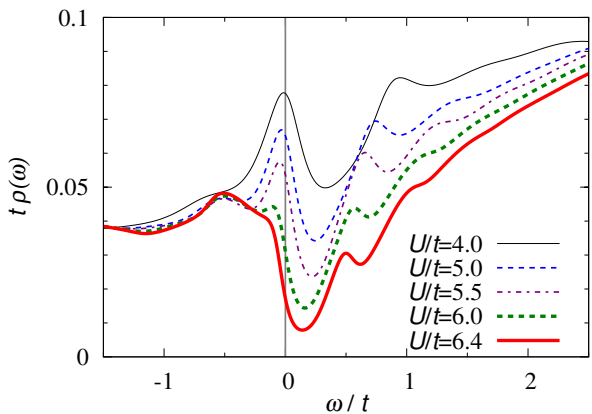}
\caption{
(Color online) 
Density of states at $T/t=0.1$ for 
several interaction strength 
$U=U'$ without the Hund's coupling $J=0$.}
\label{fig:dos_u}
\end{center}
\end{figure}

\begin{figure}[tb]
\begin{center}
\includegraphics[clip,width=\figwidth]{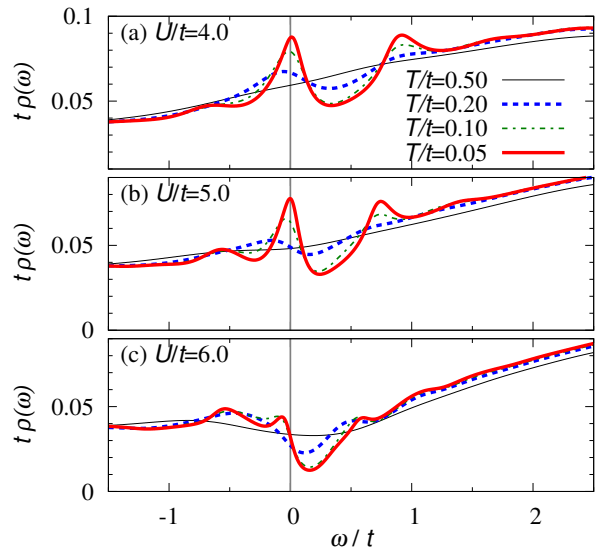}
\caption{
(Color online) 
Temperature dependence of the density of states at 
(a) $U/t=U'/t=4.0$, (b) $5.0$ and (c) $6.0$ for $J=0$.}
\label{fig:dos_t}
\end{center}
\end{figure}

Let us now investigate spatial fluctuations of spin and orbital 
in the two-orbital Hubbard model at quarter filling 
using CDMFT + NCA. 
We have confirmed that our results for $U'=J=0$, 
which corresponds to the single band Hubbard model, 
are consistent with the previous results \cite{stanescu06}. 
We first study the case without the Hund's coupling. 
In Fig. \ref{fig:dos_u}, we show the density of states (DOS) 
for $U=U'$, $J=0$ at a temperature $T/t=0.1$. 
Here, the DOS per spin and orbital is defined as 
$\rho(\omega)=-\mathrm{Im}{G_{ii \alpha \sigma}}(\omega+i0)/\pi$. 
We can clearly see the metal-insulator crossover in the DOS. 
As $U$ increases, the quasiparticle peak around 
the Fermi level becomes sharper. 
We can clearly see the heavy quasiparticle peak 
for $U/t=4.0$ in Fig. \ref{fig:dos_u}. 
As further increasing $U$, the quasiparticle peak is suppressed and 
vanishes at $U/t \sim 5.8$. 
For $U/t > 5.8$, a gap is formed near the Fermi level 
and the system becomes insulating. 
We thus find that the metal-insulator boundary exists 
at $U/t \sim 5.8$ for $T/t=0.1$. 
The $T$-dependence of the DOS is shown in Fig. \ref{fig:dos_t}. 
In the metallic phase ($U/t=4.0$), 
the quasiparticle peak appears near the Fermi level, 
as $T$ is lowered below the FL coherence temperature $T_0/t \sim 0.5$. 
Here, $T_0$ is defined as the width of 
the quasiparticle peak at low temperatures. 
Around the metal-insulator boundary ($U/t=5.0$), 
the coherence temperature becomes lower ($T_0/t \sim 0.3$) 
and the sharper peak develops at low temperatures. 
This behavior clearly demonstrates the evolution of 
the heavy quasiparticle at low temperatures. 
On the other hand, in the insulating phase for $U/t=6.0$, 
the gap becomes more prominent with lowering $T$, 
although DOS at the Fermi level $\rho(0)$ 
is finite at $T/t \ge 0.05$. 
At much lower temperatures, $\rho(0)$ should vanish and 
the system is expected to be a real insulator. 
These results are quite different from those of 
the single-band Hubbard model at half filling 
on the square lattice \cite{stanescu06,moukouri01,imai02,maier05}, 
but are consistent with those of single-site DMFT 
\cite{rozenberg97,inaba07}. 
In the former case, 
where one electron occupies each site 
as well as our model, 
the FL quasiparticle is strongly suppressed 
by antiferromagnetic (AFM) fluctuations. 
On the other hand, in the latter case, 
where spatial correlations are not taken into account, 
the quasiparticle peak appears around the Mott transition. 
Therefore, in our case, 
spatial correlations of spin and orbital are 
strongly suppressed by orbital degeneracy, 
and the FL quasiparticles are stabilized. 

\begin{figure}[tb]
\begin{center}
\includegraphics[clip,width=\figwidth]{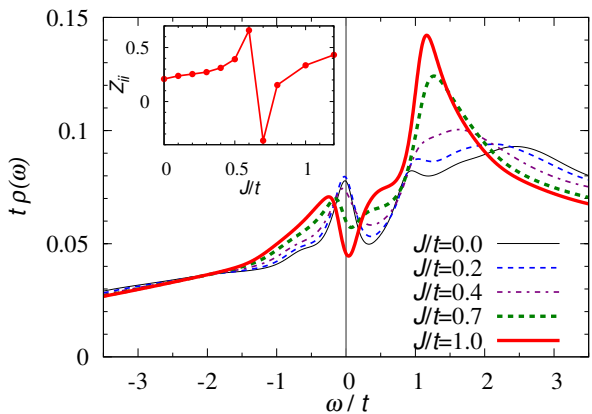}
\caption{
(Color online) 
Density of states for $U/t=4.0$ at $T/t=0.1$ 
with varying $J$. 
The inset shows the renormalization factor 
as a function of $J$.
The renormalization factor for $J/t > 0.7$ has no meaning 
because of the Fermi-liquid break-down. }
\label{fig:dos_j}
\end{center}
\end{figure}

We next investigate the system with the Hund's coupling. 
In the presence of the Hund's coupling, 
we find that the spatial correlations give rise to 
the pseudo gap in the quasiparticle excitation. 
In Fig. \ref{fig:dos_j}, we show the DOS 
for $U/t=4.0$ at $T/t=0.1$ with varying 
the Hund's coupling $J$. 
As the Hund's coupling $J$ increases, 
the quasiparticle peak gradually shrinks 
and vanishes for $J/t \sim 0.7$. 
As $J$ further increases, 
a pseudo gap evolves and the system becomes insulating. 
We note that 
this behavior caused by the Hund's coupling $J$ 
is qualitatively different from 
the results of single-site DMFT \cite{inaba07}. 
Within DMFT, which does not treat the spatial correlations, 
it is known that the Hund's coupling reduces 
the energy gap of the Mott insulator to $U'-J$. 
Therefore, in DMFT, the Hund's coupling 
reduces the effective Coulomb repulsion and 
tends to stabilize the metallic phase. 
The previous DMFT + NCA studies show that 
the quasiparticle peak in DOS get enhanced as $J$ increases \cite{imai01}. 
On the other hand, 
in our CDMFT, which properly incorporate spatial correlations, 
the Hund's coupling enhances spatial correlations 
of spin and orbital, which give rise to 
the NFL state with the pseudo gap. 

The NFL behavior is also seen 
in the renormalization factor 
$
\hat{Z}=\left[
\left .
\hat{1}-\partial \mathrm{Re}\hat{\Sigma}( \omega +i0 )/
  \partial \omega
\right|_{\omega=0}
\right]^{-1}
$. 
In the inset of Fig. \ref{fig:dos_j}, 
we show $Z_{ii}$ as a function of $J$. 
With increasing $J$, 
the renormalization factor $Z_{ii}$ increases, 
and jumps to negative values at $J/t \sim 0.7$. 
The negative values of $Z_{ii}$ indicate the break-down of FL. 
We thus find the crossover from FL to NFL states at $J/t \sim 0.7$. 
This behavior is consistent with the results of DOS in Fig \ref{fig:dos_j}. 
In single-site DMFT, the renormalization factor increases as $J$ increases, 
but its mechanism is completely different from that in our case. 
Within DMFT, the effective interaction is reduced by $J$, 
as mentioned above \cite{inaba07}. 
Therefore, the renormalization factor increases and the FL metallic 
state is stabilized. 
On the other hand, in our CDMFT, which incorporates 
not only such local renormalization effects of $J$ 
but also nonlocal correlations, 
$Z_{ii}$ increases and shows the NFL properties. 
This behavior is also seen in the dynamical cluster studies 
of the single-band Hubbard model at half-filling \cite{maier05,imai02}. 
In this case, the AFM correlations disturb the FL formation 
and induce the NFL behavior. 
In our case, the AFO correlations enhanced 
by the Hund's coupling trigger the NFL properties in $Z_{ii}$. 

\begin{figure}[tb]
\begin{center}
\includegraphics[clip,width=\figwidth]{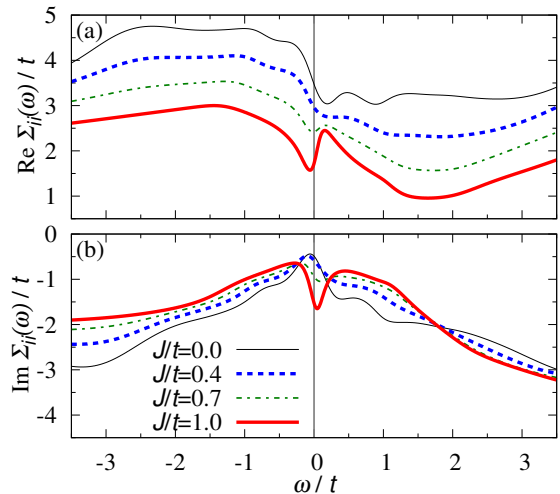}
\caption{
(Color online) 
(a) The real part and (b) the imaginary part of the 
self-energy for $U/t=4.0$ at $T/t=0.1$ 
with varying $J$. 
}
\label{fig:se_j}
\end{center}
\end{figure}

In Fig. \ref{fig:se_j}, we also show the local self-energy 
$\Sigma_{ii}( \omega +i0 )$ for typical values of $J$. 
For small $J$ ($J/t=0, 0.4$), 
$\mathrm{Im} \Sigma_{ii}( \omega + i0 )$ is small and 
$\partial \mathrm{Re} \Sigma_{ii}( \omega + i0 )/\partial \omega$ 
is negative at $\omega=0$, which is FL-like behavior. 
As $J$ increases, 
$\mathrm{Im} \Sigma_{ii} (\omega + i0)$ at $\omega = 0$ 
increases in the negative direction 
and the profile of $\mathrm{Im} \Sigma_{ii} (\omega + i0)$ 
dramatically changes in the large $J$ region ($J/t=1.0$). 
Also, $\partial \mathrm{Re} \Sigma_{ii}( \omega + i0 )/\partial \omega$ 
at $\omega = 0$ changes its sign 
from negative to positive at $J/t \sim 0.7$. 
These results are consistent with the results of DOS and 
the renormalization factor. 

\begin{figure}[tb]
\begin{center}
\includegraphics[clip,width=\figwidth]{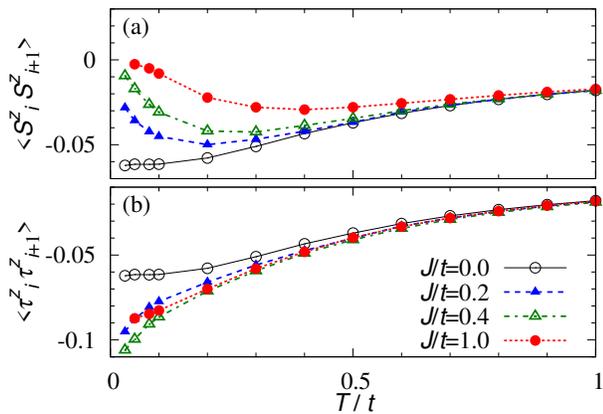}
\caption{
(Color online) 
Nearest-neighbor correlation functions of 
(a) spin $\langle S^z_i S^z_{i+1} \rangle$ and 
(b) orbital $\langle \tau^z_i \tau^z_{i+1} \rangle$, 
as a function of the temperature $T$ at $U/t=4.0$. }
\label{fig:cf}
\end{center}
\end{figure}

To clarify how the insulating state is induced by 
spatial correlations, 
we calculate the nearest-neighbor spin correlation function 
$\langle S^z_i S^z_{i+1} \rangle$ and 
orbital correlation function 
$\langle \tau^z_i \tau^z_{i+1} \rangle$. 
Here, $S_i^z$ and $\tau_i^z$ are defined as 
$S_i^z=\sum_\alpha (n_{i \alpha \uparrow}-n_{i \alpha \downarrow})/2$ and 
$\tau_i^z =\sum_\sigma (n_{i 1 \sigma}-n_{i 2 \sigma})/2$, respectively. 
In Fig. \ref{fig:cf} (a), 
we show the $T$-dependence of 
the spin correlation function 
$\langle S^z_i S^z_{i+1} \rangle$ for several values of $J$. 
For $J=0$, $\langle S^z_i S^z_{i+1} \rangle$ is 
always negative, so that the spin correlations are AFM, 
which are monotonically enhanced with lowering $T$. 
On the other hand, 
for the finite Hund's coupling $J$, 
$\langle S^z_i S^z_{i+1} \rangle$ once decreases as $T$ decreases and 
then upturns, taking a minimum at $T=T^*$. 
At much lower temperatures, 
$\langle S^z_i S^z_{i+1} \rangle$ 
tends to become positive and 
the spin correlations are expected to be ferromagnetic (FM). 
For the larger interaction $U/t=6.0$, 
we find that the spin FM correlations are more enhanced 
at low temperatures. 
This behavior indicates that 
the effects of the Hund's coupling 
become prominent at $T<T^*$, 
and the AFM correlations are strongly suppressed. 
The characteristic temperature $T^*$ increases 
with increasing $J$: 
$T^*/t\sim 0.2$, $0.3$ and $0.4$ for $J/t=0.2, 0.4$ and $1.0$, 
respectively.  
Fig. \ref{fig:cf} (b) shows the $T$-dependence of 
the orbital correlation function 
$\langle \tau^z_i \tau^z_{i+1} \rangle$. 
The orbital correlation functions 
are negative 
and the correlations are AFO. 
As $T$ decreases, the AFO correlations 
gradually become strong and 
get strongly enhanced at $T<T^*$ in the presence of the Hund's coupling. 
For $J/t=1.0$, the AFO correlations are suppressed at low temperatures, 
which is due to the suppression of 
the orbital moment by the Hund's coupling. 

The noticeable point is that 
the AFO correlations for $J=0$ at low temperatures 
are much weaker than those for finite $J$. 
For $U=U'$, spin and orbital states in two adjacent sites
are highly degenerate, {\it i.e.} 
AFM and AFO, AFM and ferro-orbital, and FM and AFO 
states are all degenerate, because of the spin and orbital symmetry. 
This degeneracy strongly suppresses 
the nearest-neighbor spin and orbital correlations. 
On the other hand, the Hund's coupling lifts the degeneracy, 
and strongly enhances the AFO fluctuations at $T<T^*$. 
These spatial correlations 
also affect the quasiparticle dynamics. 
For $J=0$, where the correlations are very weak due to the degeneracy, 
the FL state is stabilized and 
the heavy quasiparticle peak instead of the pseudo gap appears in the DOS, 
as seen in Fig. \ref{fig:dos_t}. 
On the other hand, the AFO correlations get strongly enhanced 
under influence of the Hund's coupling, 
and these correlations induce the pseudo gap behavior 
in the DOS, as shown in Fig. \ref{fig:dos_j}. 

\begin{figure}[tb]
\begin{center}
\includegraphics[clip,width=\figwidth]{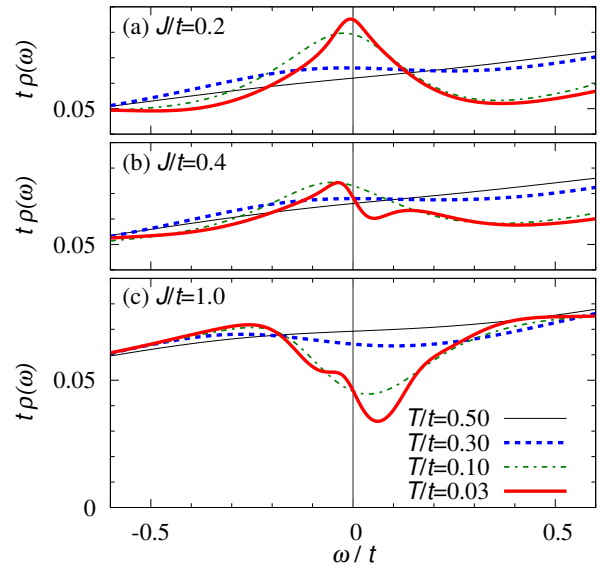}
\caption{
(Color online) 
Temperature dependence of the 
density of states at $U/t=4.0$ 
for (a) $J/t=0.2$, (b) $J/t=0.4$, and (c) $J/t=1.0$.}
\label{fig:dos_t_j}
\end{center}
\end{figure}

By investigating the $T$-dependence of the DOS 
for the different $J$, 
we find more striking behavior in the quasiparticle dynamics. 
In Fig. \ref{fig:dos_t_j}, we show the $T$-dependence of the DOS 
for $U/t=4.0$ with varying the Hund's coupling $J/t=0.2$, $0.4$ and $1.0$. 
For the weak Hund's coupling $J/t=0.2$, 
the quasiparticle peak appears at $T<T_0$, which 
becomes sharper with lowering $T$ 
as well as the case without the Hund's coupling shown in Fig. \ref{fig:dos_t}. 
In contrast, for the strong Hund's coupling $J/t=1.0$, 
the pseudo gap appears around the Fermi level, 
and the gap becomes prominent at $T<T^* \sim 0.4t$. 
For the intermediate Hund's coupling $J/t=0.4$, 
the two energy scales, $T_0$ and $T^*$, become relevant to 
the quasiparticle dynamics. 
This results in the nonmonotonic $T$-dependence of the DOS. 
For $J/t=0.4$, as $T$ decreases, 
the quasiparticle peak once develops at $T<T_0$. 
At much lower temperatures $T<T^* \sim 0.3t$, 
the AFO correlations get enhanced, 
and the DOS shows insulating behavior with a dip near the Fermi level. 
We thus conclude that the nonmonotonic $T$-dependence is 
induced by the competition between the FL formation 
and the AFO correlations. 

\section{Summary}

We have studied the effects of spatial fluctuations 
in the two-orbital Hubbard model at quarter filling 
by means of CDMFT + NCA. 
We have found the heavy quasiparticle behavior 
around the metal-insulator boundary, 
which is caused by orbital degeneracy. 
It has been clarified that the Hund's coupling 
enhances the AFO fluctuations, 
which gives rise to the pseudo gap behavior of the DOS. 
We have also found a novel nonmonotonic 
$T$-dependence in the single-particle excitations 
\cite{ohashi08} caused by the FL formation and 
the AFO correlations for the intermediate Hund's coupling. 

It has been suggested that the ground state of 
the two-orbital Hubbard model at quarter filling is 
FM and AFO ordered state in strong coupling region \cite{momoi98,kubo09}. 
Also in our study, we have found that the AFO correlations 
get enhanced by spatial fluctuations due to the Hund's coupling, 
which is naturally expected to stabilize 
the ordered phase at zero temperature. 
In the two dimensional $SU(4)$ spin-orbital model, 
the importance of the plaquette singlet correlation 
has been pointed out \cite{li98}. 
In our CDMFT, however, we have used the two-site effective cluster model 
as a minimal model to study the effects of the spatial fluctuations and 
have not incorporate the spin and orbital fluctuations in a plaquette. 
Therefore, the effects of the correlations in the plaquette should 
be studied using the larger cluster CDMFT. 
Also, to quantitatively improve our results, 
we should use the essentially exact cluster-solver, 
such as the continuous-time quantum Monte Carlo method, 
in our future work.  
On the other hand, 
in the real materials with the inter-layer hopping, 
the correlations in a plaquette are not strong, but
the AFO correlations are expected to be dominant. 
Therefore, we expect that our findings in the present study, 
such as the heavy FL behavior 
and the nonmonotonic $T$-dependence 
in quasiparticle excitations, will be observed experimentally 
in the correlated electron systems with orbital degeneracy. 

\section*{Acknowledgments}
The authors thank H. Tsunetsugu, N. Kawakami, T. Momoi and K. Inaba 
for valuable discussions. 
This work has been supported by 
the Japan Society for the Promotion of Science, 
Grant-in-Aid for Scientific Research (No. 21740232, No. 20540390), 
the Next Generation Supercomputing Project, Nanoscience Program 
from the Ministry of Education, Culture, 
Sports, Science and Technology, Japan. 
A part of the computations was done at the Supercomputer
Center at the Institute for Solid State Physics, University of Tokyo and 
Yukawa Institute Computer Facility.

\end{document}